\documentclass[11pt]{article}

\usepackage{cite}
\usepackage{graphicx}
\author{Stefan Janiszewski and Andreas Karch}
\title{Moving Defects in AdS/CFT}
\begin{document}

\maketitle
\begin{center}\emph{Department of Physics, University of Washington, Seattle, WA 98195}\end{center}
\begin{abstract}We study defects of various dimensions moving through Anti-de Sitter space. Using the AdS/CFT correspondence this allows us to probe aspects of the dual quantum field theory. We focus on the energy loss experienced by these defects as they move through the CFT plasma. 
We find that the behavior of these physical quantities is governed
by induced world-volume horizons. We identify world-volume analogs
for several gravitational phenomena including black holes,
the Hawking-Page phase transition and expanding cosmological horizons.  \end{abstract}
\newpage

\section{Introduction}
The $AdS/CFT$ correspondence \cite{JM,EW,GKP} has allowed the exploration of strongly coupled conformal field theories via gravitational calculation in their dual (classical) $AdS$ spaces. This correspondence, like most dualities, is very powerful in allowing exploration of intractable regimes of one theory via computation in the dual theory. Not only can we learn properties of strongly coupled field theories by studying gravitational spaces, the correspondence also gives a non-perturbative definition of quantum gravity.

Of interest for this paper is the study of world-volume defects in the $AdS$ space, and the extraction of properties of the $CFT$ via the behavior of the boundary values of these defects. We will closely follow the methods of \cite{HKKKY,Gubser:2006bz,CST,LRW,Liu:2006he,SSG,BWX} where a probe string defect is added to the $AdS_5$ and the properties of the endpoints on the boundary give information on the behavior of point objects in the dual $\mathcal{N} = 4$ $SYM$ plasma, such as drag and screening length. We will continue this line of reasoning by considering extended defects of various dimensions in $AdS_{d+1}$. Related studies of moving \cite{FM} and rotating \cite{takayanagi} extended objects, also governed by world-volume black holes, have appeared recently.

Most of the focus of the paper will be on membrane-like defects in $AdS_7$. The reasons for this are various. It could be expected that the extended spatial nature of these defects on the boundary, as compared to the quark-like endpoints of previously studied strings, will probe unexplored regimes of the hydrodynamic description of the $CFT$ plasma, such as turbulence. The use of $AdS_7$ is two-fold: as will be shown, the equation of motion for a moving membrane in this space enjoys analytically special properties, being in a class also containing the well studied string in $AdS_5$; secondly, $AdS_7/CFT_6$ is an example of the correspondence that is relatively unexplored and promises to be rich. Unlike the traditional usage, where four dimensional $\mathcal{N} =4$ $SU(N)$ $SYM$ is dual to ten dimensional Type II B string theory compactified on an $AdS_5$ background, the six dimensional supersymmetric $CFT_6$ and eleven dimensional M-Theory compactified on $AdS_7$ are dual descriptions of a stack of $N$ $M5$ branes.  This allows testing of the correspondence between theories of different structure than normally considered. The $CFT_6$ is a semi-mysterious field theory, it contains chiral matter, has the supersymmetry $\mathcal{N}=(0,2)$, and, unlike its four dimensional $SYM$ cousin, it currently has no known Lagrangian description. 
The theory is an isolated fixed point of the renormalization group, and therefore has no parameters, not even dimensionless ones \cite{NS}, other than the number $N$ of $M5$ branes. Additionally, $AdS/CFT$ in this case gives us a definition of M-Theory in terms of $CFT_6$ degrees of freedom (above uncertainties aside). 
In the tractable regime where we perform calculations we take the large $N$ limit of the number of $M5$ branes, which corresponds to considering the eleven dimensional supergravity limit of M-Theory. Both sides of this correspondence deserve further study and we hope that this is a step in that direction.

The paper is organized as follows: The remainder of this section is dedicated to establishing the notation and metric used to describe the geometry, and introduces the action for the defect, which will determine its equations of motion. In Section 2 we study defects that are infinitely extended and moving with a constant velocity. The effect of this motion through the warped geometry is to induce a horizon on the world-volume of the defect, a phenomena that will become a major theme of the paper. The effect of this horizon is that its temperature will feature in various physical quantities, such as the energy loss experienced by the moving defect, or the screening length between two such defects. In Section 3 we turn to defects which are compact on the boundary, and in particular are n-spheres. Here too a world-volume horizon is induced and plays a defining role: defects can truncate after extending into the bulk for a finite distance only if they do not cross the location of the induced horizon, otherwise they must extend all the way to the bulk horizon. A transition between these two classes is explored. Lastly, in Section 4 we study defects that are accelerating in pure $AdS$. Even though there is no bulk black hole, we again find a world-volume horizon due to the motion and expect its temperature to be physically relevant. Section 5 summarizes our conclusions and states some possible further studies.

\subsection{The Metric}
We will be working in $AdS_{d+1}\otimes W$, where $W$ is a compact manifold, whose exact form is dictated by symmetries of the underlying theory. Examples include the familiar $AdS_5\otimes S_5$, where $W$ is a five-sphere with radius equal to the $AdS$ radius of curvature, this being determined by the supersymmetry of II B string theory. For $d=6$, we see that $W$ is a four-sphere with radius equal to half of the radius of curvature of $AdS_7$ \cite{AGMOO}, as determined by 11 dimensional supergravity. 

We will often include a black brane with horizon located at $r_h$, giving the near-horizon metric $G_{\mu \nu}$:
\begin{equation}\label{eq:metric1}
ds^2 = R^2\left(r^2\left(-h(r) dt^2  +\sum_{i=1}^{d-1}dx_i^2 \right)+\frac{dr^2}{r^2h(r)}+R^{-2}ds_W^2\right)\quad,\end{equation} 
where: $h(r)=1-\left(\frac{r_h}{r}\right)^d$; R is the radius of curvature of the $AdS$ space; $ds_W^2$ is the line element on $W$; and $r_h$ relates to the Hawking temperature as $T=\frac{d}{4 \pi}r_h$. When working at non-zero temperature we will most often use the change of coordinates: $u\equiv\frac{r_h}{r}$, $t\to\frac{d}{4\pi T}t$, and $x_i\to\frac{d}{4\pi T}x_i$. This has the horizon at $u=1$, boundary at $u=0$, and gives the metric:
\begin{equation}\label{eq:metric2}
ds^2 = \left(\frac{R}{u}\right)^2 \left(-\left(1-u^d\right)dt^2+\frac{du^2}{\left(1-u^d\right)}+d\vec{x}^2+\left(\frac{u}{R}\right)^2ds_W^2 \right)\quad.
\end{equation}

Additionally, we will often consider truncating the asymptotic $AdS$ space down to a maximal bulk radial value $r=r_m$. This will regulate the behavior near the boundary. Ending the profile of a bulk defect at this radius corresponds to a defect with finite rest mass density in the $CFT$, as the asymptotically filling D-brane in \cite{HKKKY} lead to a finite rest mass of the dual quarks.

\subsection{The Action}
We will be interested in extended objects in the $CFT$ which appropriately ``hang'' in the radial direction from $r=r_m$ towards the horizon. In the strong coupling, large $N$ regime of the $CFT$, the bulk theory will be given by classical gravity. We can then use a Nambu-Goto-like action to determine the behavior of the defect. For the world-volume map $\sigma^i\mapsto X^\mu$
the classical equations of motion for the defect will be given by variation of the action:
\begin{equation}\label{eq:action}S=-T_0\int \prod_i d\sigma_i \sqrt{-g}\quad, \end{equation}
where $T_0$ is the tension of the defect, and $g \equiv det\left(g_{ab}\right)$ for the world-volume metric $g_{ab}$ induced from the bulk $AdS$ space (\ref{eq:metric1}). Denoting partial differentiation with respect to world-volume coordinates as $\frac{\partial A}{\partial b} \equiv A_{,b}$, we simply have $g_{ab} = G_{\mu \nu}X^\mu_{,a} X^\nu_{,b}$.

The dimensionless parameter $T_0 R^{n+2}$ will play the role of quantifying the classical as well as the probe limits (``large $N$, strong coupling'' limits in the field theory). We will require $T_0 R^{n+2}\gg 1$ so that the action is large 
and the classical solution dominates, allowing us to ignore quantum corrections in the gravitational theory. 
Furthermore, we will work in the probe limit 
$T_0 R^{n+2}\ll (M_{Pl,d+1} R)^{d-1}$ where $M_{Pl,d+1}$ is the $d+1$-dimensional Planck mass. This allows us to ignore the gravitational back reaction of the defect on the $AdS$ geometry.
For example, a string in $AdS_5$ has $T_0 R^2=\sqrt{\lambda}/2\pi$ where $\lambda =g^2_{YM}N$ is the 't Hooft coupling of the $SYM$. 

\section{Infinitely Extended Defects}
\subsection{Equation of Motion}
We will first consider coordinate parameterizations of the form applicable to an object infinitely extended in $n$ transverse directions, $\vec{y}$, and moving in an additional one of the transverse directions, $x$. In static gauge the world-volume map is therefore of the form $X = \left(t,r,\vec{y}, x\left(t,r,\vec{y}\right),\vec{z}=const,W=const \right)$ where $\vec{z}$ are the additional transverse coordinates on which the world-volume does not depend. This map gives the determinant of the induced world-volume metric:
\begin{equation}\label{eq:g1} -g = R^{2\left(n+2\right)}r^{2n}\left(1+\sum_{i=1}^n x^2_{,y_i}+r^4h(r)x^2_{,r} -\frac{x^2_{,t}}{h(r)}\right)\quad. \end{equation} 

Variation of the action (\ref{eq:action}) with respect to the function $x$ gives the equation of motion:
\begin{equation}\label{eq:eqmo}
\left(\frac{x_{,t}}{\sqrt{-g}}\right)_{,t}=\frac{h}{r^{2n}}\left(\frac{r^{2\left(n+2\right)}hx_{,r}}{\sqrt{-g}}\right)_{,r}+h\sum_{i=1}^n\left(\frac{x_{,y_i}}{\sqrt{-g}}\right)_{,y_i}\quad.
\end{equation}
We can also vary the Lagrangian corresponding to (\ref{eq:action}) to find the momentum densities:
\begin{equation}\label{eq:den} \Pi_\mu^{\left(\sigma_i\right)} \equiv \frac{\delta \mathcal{L}}{\delta X^\mu_{,\sigma_i}} = -T_0 \sqrt{-g}g^{\sigma_i a}G_{\mu \nu}X^\nu_{,a}\quad.
\end{equation}

\subsection{Solutions}
\subsubsection{Static Defect}
A static object hanging straight down, $x(t,r,\vec{y}) = x_0$, trivially satisfies the equations of motion (\ref{eq:eqmo}) and from (\ref{eq:den}) has energy density:
\begin{displaymath} -\Pi^{\left(t\right)}_t = \frac{T_0 R^{2\left(n+2\right)} r^{2n}}{\sqrt{-g}} \left(1+\sum_{i=1}^n x^2_{,y_i} + r^4h x^2_{,r}\right) = T_0 R^{n+2}r^n\quad. \end{displaymath}
Integrating this over the radial extent gives the energy per unit volume in the $\vec{y}$ space:
\begin{displaymath} \epsilon = T_0 R^{n+2} \int_{r_h}^{r_m} dr r^n =\frac{T_0 R^{n+2}}{n+1}\left(r_m^{n+1}-r_h^{n+1}\right)\quad.\end{displaymath}
Following the $AdS/CFT$ ``dictionary,'' in the zero temperature limit corresponding to $r_h = \frac{4 \pi T}{d} = 0$, this energy should be the Lagrangian mass per unit volume, $\mu$, of the n-dimensional object in the $CFT$. Therefore we have: 
\begin{displaymath} \mu = \frac{T_0 R^{n+2}}{n+1}r_m^{n+1}\quad.\end{displaymath}

\subsubsection{Uniformly Moving Object}
For this section we find it useful to work with the form (\ref{eq:metric2}) of the metric. We now consider a defect that has a stationary profile, but moves with a constant ``velocity.'' For the ansatz of an object whose shape only depends on the radial coordinate $u$, and has a constant time derivative, $v$, we have the parametrization:
\begin{displaymath} x(t,u,\vec{y}) =v t+x(u)\quad. \end{displaymath}
This leaves $-g$ time independent and the equation of motion (\ref{eq:eqmo}) reduces to:
\begin{displaymath}  
\left(\frac{h x_{,u}}{u^{2\left(n+2\right)}\sqrt{-g}}\right)_{,u} = 0\quad.\end{displaymath}
The quantity in parenthesis is therefore a constant, $C v / R^{n+2}$. Solving this equation for $x_{,u}^2$ we get:
\begin{displaymath} x_{,u}^2 = \left(\frac{C v u^{n+2}}{h}\right)^2 \frac{1-h^{-1}v^2}{1-C^2 v^2 h^{-1}u^{2\left(n+2\right)}}\quad. \end{displaymath}

We see that both the numerator and denominator  have a zero and change sign as $u$ varies. For both $x_{,u}^2$ and $-g$ to be positive at all points we require this sign change to happen at the same radial coordinate, $u_c$, for both terms in the fraction. This yields:
\begin{displaymath} u_c = \left(1 - v^2\right)^\frac{1}{d}=\gamma^{-2/d} \quad, \end{displaymath}
\begin{displaymath} C = \pm 1/u_c^{n+2} = \pm \gamma^\frac{2\left(n+2\right)}{d}\quad, \end{displaymath}
where $\gamma$ is the Lorentz factor related to $v$.

\begin{figure}[!htb]
	\centering
	\includegraphics[width=0.62\textwidth]{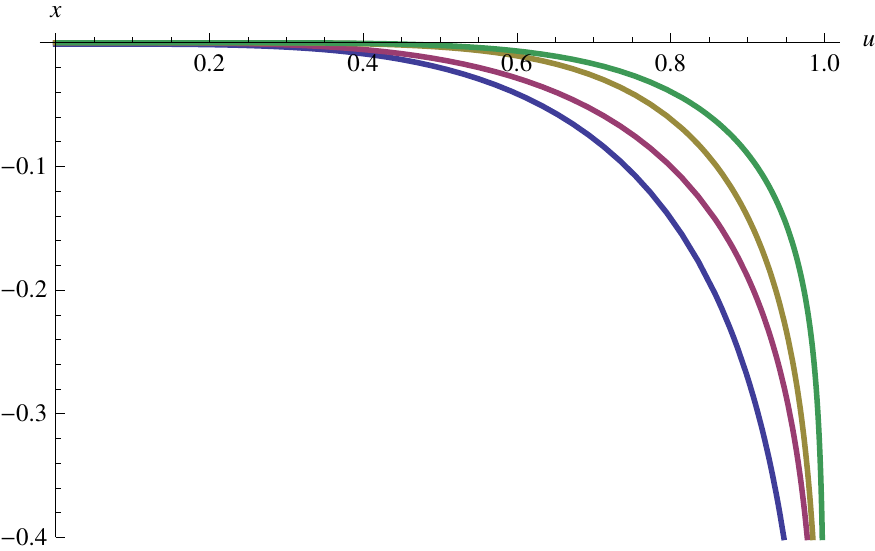}
	\caption{Profiles of uniformly moving defects.
From left to right- n=1, d=4; n=1, d=6; n=3, d=6; n=3, d=10. All have $\gamma=6$ and $C<0$. $C>0$ reflects across the $u$ axis. (Color online)}
	\label{fig:fig1}
\end{figure}

We now have an ordinary first-order differential equation for $x(u)$ that we can solve for various values of $d$ and $n$. Generically, the solutions have the qualitatively similar behavior in that they hang down from the asymptotic radial value and sweep an arc down to a tangential approach of the horizon. Figure (\ref{fig:fig1}) illustrates some examples. 

From an analytic standpoint there is a class of special objects based on their dimensionality and the dimension of the bulk. We see that for the situation of $d=2(n+2)$ the equation for $x_{,u}$ reduces greatly to:
\begin{displaymath}
x_{,u}=\pm v \frac{u^{d/2}}{1-u^d}\quad,\end{displaymath}
which has the solution:
\begin{equation}\label{eq:xspec} x(u)=\pm\frac{v}{d} B\left(u^d,\frac{1}{2}+\frac{1}{d},0\right)\quad,\end{equation}
where $B$ is the incomplete Euler beta function. For $n=0$ and $d=4$ this is the well studied case of a string in $AdS_5$, with solution:
\begin{displaymath}x=\pm\frac{v}{2}\left(\tanh^{-1}(u)-\tan^{-1}(u)\right)\quad.\end{displaymath}

While for $n=1$ we have a membrane-like object in $AdS_7$, dual to a one dimensional defect in the corresponding supersymmetric $CFT_6$, with the profile:
\begin{displaymath} x=\pm\frac{v}{12}\left(2\sqrt{3}\left(\tan^{-1}\left(\frac{2u+1}{\sqrt{3}}\right)-\tan^{-1}\left(\frac{2u-1}{\sqrt{3}}\right)\right)+\log\left(\frac{u^4+u^2+1}{\left(1-u^2\right)^2}\right)\right)\quad.\end{displaymath} 

\subsubsection{Energy Loss}
In general there are still two possible solutions, depending on the sign of $C$. This can be fixed by physical criteria. Due to the nature of the black brane, we expect energy and momenta to only flow away through the horizon at $u=1$, and not into the bulk, towards smaller values of $u$ and the boundary. This physical representation is found by examining the momentum currents:
\begin{displaymath} 
\Pi^{\left(u\right)}_t = \frac{T_0 h}{\sqrt{-g}}\left(\frac{R}{u}\right)^{2\left(n+2\right)} x_{,t}x_{,u} = T_0 R^{n+2} v^2 C \quad,
\end{displaymath}
\begin{equation}
\Pi^{\left(u\right)}_x = -\frac{T_0 h}{\sqrt{-g}}\left(\frac{R}{u}\right)^{2\left(n+2\right)} x_{,u} = -T_0 R^{n+2} v C \quad.
\end{equation}

The rate at which energy flows in the radial direction on the world-volume is given by $-\Pi^{\left(u\right)}_t$. Physically, we require this rate to be positive, i.e. energy is flowing toward larger radial coordinate $u$, that is, towards the horizon. This requires the choice $C<0$. As seen in Figure (\ref{fig:fig1}), this corresponds to the tail of the profile ``dragging'' behind the endpoint on the boundary. 

The fact that this profile experiences energy loss means that to maintain the stationary profile moving at $v$ an external force must be doing work. Therefore it should be understood that properly there is a boundary force term in the defect action, with the field strength needed to move the defect steadily along. 

Restoring units to our coordinates, via $r_h=4\pi T/d$, the rate of energy loss becomes:
\begin{displaymath} 
-\Pi^{\left(u\right)}_t  = T_0 R^{n+2} v^2\gamma^\frac{2(n+2)}{d} \left(\frac{4\pi T}{d}\right)^{n+2}= T_0 R^{n+2}\left(\frac{4\pi}{d}\right)^{n+2} v^2 T_{eff}^{n+2}\quad,
\end{displaymath}
where we have defined $T_{eff}\equiv \gamma^{2/d}T$. 

This velocity dependence of the effective temperature felt by the probe defect can be interpreted on both sides of the duality. The energy density of the $CFT$ plasma, due to conformal invariance, must scale like $\epsilon\propto T^d$. Being the zero-zero component of the stress-energy tensor, the moving defect will see this density Lorentz transformed, picking up a blue shift factor of $\gamma^2$. This tells us the probe interacts with a plasma of effective energy density $\epsilon_{eff}\propto \gamma^2T^d=T_{eff}^d$, and therefore all loss rates only ``see'' the effective blue shifted temperature $T_{eff}$. This can be taken as an indication that the hydrodynamic description of the $CFT$ plasma is very precise: a fast probe only interacts with the blue shifted hydrodynamic variable $\epsilon_{eff}$ and does not see any other structure of the field theory. This is the picture originally advertised for the case of a string in \cite{LRW}. Here we find strong additional support by showing that this
simple blueshift of the energy density accounts for the velocity dependence for any $n$ and $d$. We would expect $1/N$ corrections to effect this picture, as these will alter the stress-energy tensor.

On the bulk $AdS$ side we can also understand the dependence of defect quantities on $T_{eff}$. As we will see in the next section, at the radial coordinate $u_c$ there is a horizon on the world-volume of the defect. From the view of the defect near the boundary the bulk horizon at $u=1$ is hidden behind the world-volume horizon. Any thermal behavior of the defect should therefore be determined by the temperature of the horizon at $u_c$ which is easily seen to correspond to $T/u_c=\gamma^{2/d}T\equiv T_{eff}$. Here too we expect there to be sub-leading corrections, in this case quantum gravity corrections to the classical $AdS$ description of the bulk. For example, a graviton can propagate from near the bulk horizon and interact with the defect near the boundary, leaving some imprint of the temperature $T$. 

In cases where the $AdS/CFT$ dictionary is sufficiently developed we can translate these loss rates into parameters of the corresponding $CFT$. For $d=4$ and $n=0$ we recover the familiar results of a heavy quark in $\mathcal{N} =4$  $SU(N)$ $SYM$ \cite{HKKKY}. For the case of $d=6$ and $n=1$ we have the relation between the tension of an $M2$-brane and the number of $M5$-branes in the stack, $R^3 T_0=2N/\pi$ \cite{AGMOO}. In this case we can translate the rates of energy and momentum flow towards the horizon as: 
\begin{eqnarray} -\Pi^{\left(u\right)}_t = \frac{16\pi^2}{27}N\frac{v^2}{\sqrt{1-v^2}}T^3\quad,\nonumber \\
\Pi^{\left(u\right)}_x = \frac{16\pi^2}{27}N\frac{v}{\sqrt{1-v^2}}T^3\quad.
\end{eqnarray}

\subsubsection{World-Volume Horizon}

We return to the form of $x_{,u}$ and a discussion of the physical interpretation of $u_c$. As discussed by \cite{SSG} and \cite{CST} at this radial coordinate a horizon develops on the world-volume. This can easily be seen by diagonalizing the metric $g_{ab}$. From:
\begin{displaymath} ds^2_{wv} = g_{tt}dt^2+2g_{tu}dtdu+g_{uu}du^2+g_{yy}d\vec{y}^2\quad, \end{displaymath}
we see that for the reparametrization:
\begin{displaymath} \hat{u} \equiv \frac{u}{u_c},\quad\hat{t}\equiv u_c^{\frac{d-2}{2}}\left(t+f(u)\right),\quad\hat{y}\equiv\frac{y}{u_c}\quad,\end{displaymath}
where:
\begin{displaymath} f'(u)=\frac{g_{tu}}{g_{tt}}=-\frac{v x'(u)}{u_c^d-u^d}\quad,\end{displaymath}
we have the diagonal metric:
\begin{equation}\label{eq:diagmetric} ds^2_{wv}=\left(\frac{R}{\hat{u}}\right)^2\left(-\left(1-\hat{u}^d\right)d\hat{t}^2+\frac{d\hat{u}^2}{1-\left(u_c \hat{u}\right)^d-\left(1-u_c^d\right)\hat{u}^{2\left(n+2\right)}}+d\hat{y}^2\right)\quad.\end{equation}

In this form it is apparent that there is a horizon on the world-volume at the radial distance $\hat{u}=1$, or $u=u_c=\gamma^{-\frac{2}{d}}$, as here $g_{\hat{t}\hat{t}}$ vanishes. This illuminates the physical determination of the integration constant $C$. It is required to be such that when $g_{\hat{t}\hat{t}}$ changes sign so does $g_{\hat{u}\hat{u}}$ so that the signature of the induced metric remains fixed.

Furthermore, it is apparent what makes the class of $d=2(n+2)$ mentioned above ``special.'' For these dimensionalities the radial component of the metric in the hatted coordinates becomes $g_{\hat{u}\hat{u}}=\left(\frac{R}{\hat{u}}\right)^2\frac{1}{1-\hat{u}^d}$ and its form is $AdS$-like, most notably the metric is velocity independent. This class includes the well studied string-like $n=0$ object in $AdS_5$, and the more mysterious membrane-like $n=1$ object in $AdS_7$.

\subsubsection{Two Uniformly Moving Profiles}
We can trivially obtain more solutions by adding together multiple solutions of the form of the previous sections, separated in a transverse direction $z$. For illustration we will consider two of the above uniformly moving ``sheets.'' They will move with equal speed in the $x$ direction, and their infinite extent on the boundary will be spanned by $\vec{y}$, as above. We will consider the case where they are uniformly separated in an additional transverse direction $z$. To explore the possibility of novel configurations obeying these boundary conditions, and possible transitions with the trivial solution, we rederive the equations of motion for the slightly generalized embedding, $X^{\mu}=(t,u,\vec{y},x=vt+x(u),z(u),const.)$ giving:
\begin{eqnarray}
\left(\frac{h x_{,u}}{u^{2(n+2)}\sqrt{-g}}\right)_{,u}=0 \quad,\nonumber\\
\left(\frac{(h-v^2) z_{,u}}{u^{2(n+2)}\sqrt{-g}}\right)_{,u}=0\quad.\nonumber
\end{eqnarray}

These equations are coupled through the factor $\sqrt{-g}$. As discussed above, the existence of a world-volume horizon is observed and requiring that the induced metric maintains signature helps to fix the integration constants. If the world-volume extends past the critical radius $u_c=\gamma^{-2/d}$ this constraint demands that $z_{,u}=0$ and $x(u)$ is the same as above, giving the solution for these boundary conditions we deem ``trivial'': two independent sheets separated in the coordinate $z$.

\begin{figure}[!htb]
	\centering
	\includegraphics[width=0.5\textwidth]{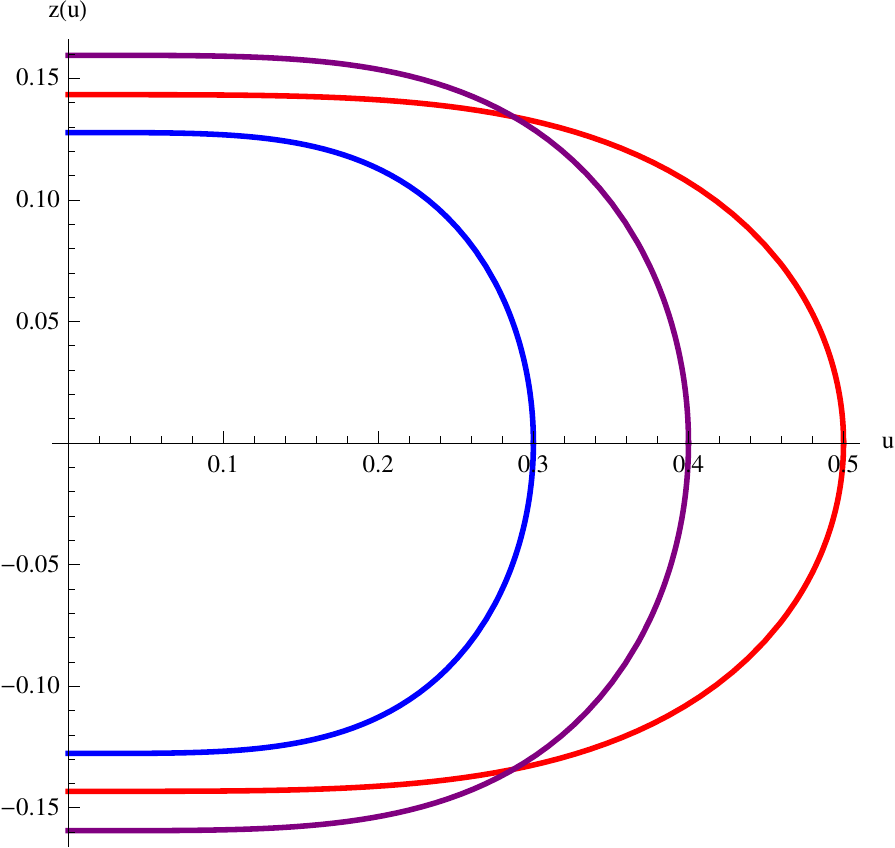}
		\caption{The connected $z$ profile for
two uniformly moving defects. The case of n=1, d=6, $\gamma=6$, and three $u_0$ are shown. Note that $u_c=6^{-1/3}\approx0.55$.} 
	\label{fig:fig2}
\end{figure}

Non-trivial solutions are found by examining the class of solutions that do not extend past the critical radius $u_c$. In this case there must be some $u_0<u_c$ which is the furthest into the bulk that the defect reaches. Examining the momentum fluxes for these general configurations we see that we require $x_{,u}=0$, otherwise there is a constant $x$ momentum flux down the world-volume that has no where to go upon reaching $u_0$. This in turn determines the other integration constant, as it is required that the $z$ profile must ``cap off'' smoothly with $z(u_0)=z(0)/2$ and $z_{,u}|_{u_0} \to -\infty$. For the special dimensionalities, $d=2(n+2)$, $z(u)$ is analytically solvable in terms of Appell hypergeometric functions. Some profiles are plotted in Figure (\ref{fig:fig2}).

\begin{figure}[!htb]
	\centering
	\includegraphics[width=0.62\textwidth]{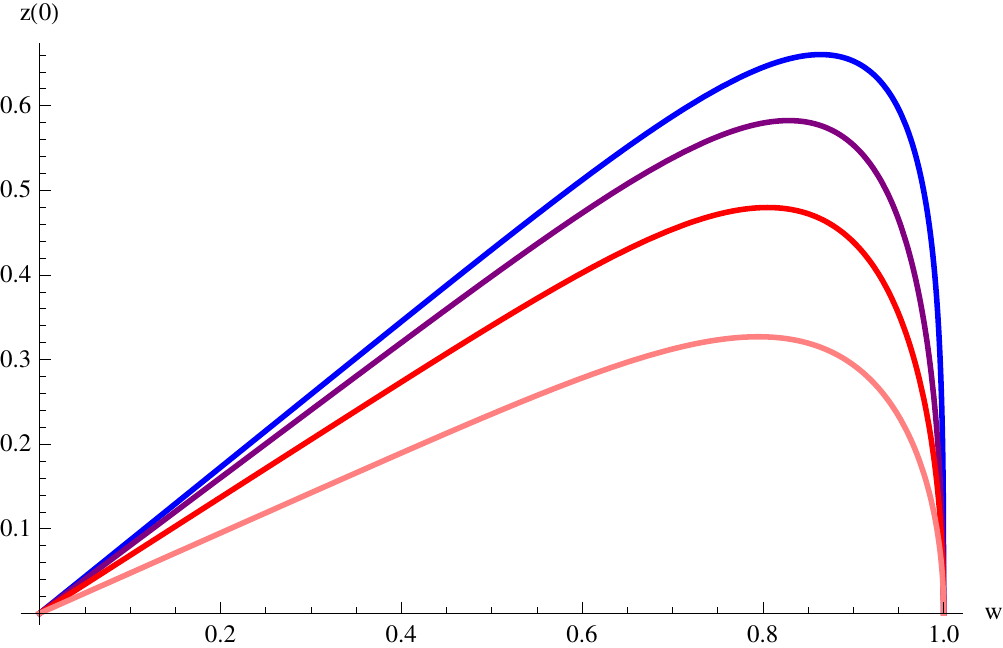}
	\caption{Boundary separation of two uniformly moving defects for n=1, d=6. Top to bottom have $\gamma=$ 1, 5/4, 2, and 6, respectively. }
	\label{fig:fig3}
\end{figure}

Of interest is the boundary separation, $z(0)$, and its dependence on the world-volume's bulk extent, $u_0$. Making the change of variables $w\equiv u_0\gamma^{2/d}$, so that $w$ can range from 0 to 1, we see that the boundary separation is given by:
\begin{displaymath}
z_w(0)=\frac{4\sqrt{\pi}\Gamma\left(\frac{3}{2}+\frac{1}{d}\right)}{(d+2)\Gamma\left(1+\frac{1}{d}\right)}w\sqrt{1-w^d}\gamma^{-2/d}{}_2F_1\left[\frac{1}{2},\frac{1}{2}+\frac{1}{d},1+\frac{1}{d},\gamma^{-2}w^d\right]\quad,
\end{displaymath}
where ${}_2F_1$ is the hypergeometric function. As apparent from Figure (\ref{fig:fig3}), plotted for various $\gamma$, this separation obtains a maximum, $z_{max}$, at some intermediate value of $w$. This is interpreted, as in \cite{LRW} and \cite{CGG}, as a screening length for the defects in the $CFT$; these connected configurations are bound meson-like states of the boundary defects, but when separated by a distance greater than $z_{max}$ the plasma effectively screens their mutual attraction and we are left with the two disconnected defects of the trivial case.

In the large velocity limit we can extract the scaling of $z_{max}$ with $\gamma$; since ${}_2F_1\to 1$ as its last argument goes to zero we see that:
\begin{displaymath}
 z_{max} \to \frac{4\sqrt{\pi}\Gamma\left(\frac{3}{2}+\frac{1}{d}\right)}{(d+2)\Gamma\left(1+\frac{1}{d}\right)}\left(1+\frac{d}{2}\right)^{-1/d}\sqrt{1-\left(1+\frac{d}{2}\right)^{-1}}\gamma^{-2/d}\quad.
 \end{displaymath}
Restoring units; $Z_{max}\equiv\frac{d}{4\pi T}z_{max}$, we see that the screening length scales as $Z_{max} \propto 1/\epsilon_{eff}^{1/d}$ in the large $\gamma$ limit. Here $\epsilon_{eff}\equiv\gamma^2\epsilon\propto T_{eff}^d$ is the blue-shifted energy density of the plasma seen by the defect. In fact, as ${}_2F_1$ weakly depends on an argument that ranges from 0 to 1, we see that this velocity scaling closely holds even in the small $\gamma$ regime, as noted in \cite{LRW} for the $d=4$ case. 

\section{Compact Objects}
We will now study compact defects. We take the spatial shape of the defect at the boundary to be an $n$-sphere, with radius $r(0)\equiv r_0$ and parametrized by azimuthal angle $\phi$, and $n-1$ polar angles $\theta_i$. Furthermore, we will consider motion of the defect in a transverse direction, $x$, which is perpendicular to its extent on the boundary. For example, for $n=1$, we have a circle on the boundary, and it will be moving in the direction parallel to the area element of the enclosed disk. Because of the symmetric nature of this motion we expect the profiles of $r$ and $x$ in the bulk to be independent of the angles. For these assumptions we can parametrize the world-volume as $(t,u,\phi,\theta_1,\dots,\theta_{n-1})\mapsto(t,u,\phi,\theta_1,\dots,\theta_{n-1},x(t,u),r(t,u),const)$. 

\subsection{Equations of Motion}
Looking for a solution similar to the uniformly moving object above we make the ansatz: $x(t,u)=vt+x(u)$ and $r(t,u)=r(u)$, for which the determinant of the induced metric becomes:
\begin{equation}\label{eq:cylmetric}-g=\left(\frac{R}{u}\right)^{2\left(n+2\right)}r^{2n}A^2\left(\left(\gamma^{-2}-u^d\right)\left(h^{-1}+r^2_{,u}\right)+hx^2_{,u}\right)\quad,\end{equation}
where $A^2\equiv \sin^{2\left(n-1\right)}(\theta_1)\sin^{2\left(n-2\right)}(\theta_2)\cdots \sin^2(\theta_{n-1})$, coming from the product of the angular parts of the metric. 

From this we obtain the equations of motion:
\begin{eqnarray}\label{eq:eomx}
\left(\frac{r^{2n}hx_{,u}}{u^{2\left(n+2\right)}\sqrt{-g}}\right)_{,u}=0\quad, \\ \label{eq:eomr} 
\left(\frac{r_{,u}r^{2n}\left(\gamma^{-2}-u^d\right)}{u^{2\left(n+2\right)}\sqrt{-g}}\right)_{,u}=\frac{nr^{2n-1}\left(\left(\gamma^{-2}-u^d\right)\left(h^{-1}+r^2_{,u}\right)+hx^2_{,u}\right)}{u^{2\left(n+2\right)}\sqrt{-g}} \quad.\end{eqnarray}

The equation (\ref{eq:eomx}) is very similar to above, we see the product in the parenthesis must be a constant, $K v/R^{n+2}$. Solving for $x^2_{,u}$ we have:
\begin{displaymath} x^2_{,u}=\left(\frac{Kvu^{n+2}}{h}\right)^2\frac{\left(\gamma^{-2}-u^d\right)\left(h^{-1}+r^2_{,u}\right)}{r^{2n}-K^2v^2h^{-1}u^{2\left(n+2\right)}} \quad.\end{displaymath}

As above, the numerator has a zero and changes sign at $u_c=\gamma^{-\frac{2}{d}}$. Likewise, for $x^2_{,u}$ and $-g$ to be everywhere positive we require the denominator to change sign at this point, and have no additional zeros. This determines $K^2=r_c^{2n}\gamma^\frac{4\left(n+2\right)}{d}$ where $r_c\equiv r(u_c)$ is the radius of the defect at the critical bulk coordinate $u_c$. We can substitute this form for $x^2_{,u}$ into the metric and obtain:
\begin{equation}\label{eq:detg} -g=\left(\frac{R}{u}\right)^{2\left(n+2\right)}r^{4n}A^2\frac{\left(\gamma^{-2}-u^d\right)\left(1+hr^2_{,u}\right)}{hr^{2n}-r^{2n}_c\left(1-\gamma^{-2}\right)\gamma^\frac{4\left(n+2\right)}{d}u^{2\left(n+2\right)}}\quad.\end{equation}

\subsubsection{World-Volume Horizon}
By the same change of coordinates above, we can again diagonalize the induced world-volume metric, obtaining the form: 
\begin{displaymath}g_{\hat{t}\hat{t}}=-\left(\frac{R}{\hat{u}}\right)^2\left(1-\hat{u}^d\right),\quad g_{\hat{u}\hat{u}}=\left(\frac{R}{\hat{u}}\right)^2\frac{r^{2n}\left(1+hr^2_{,u}\right)}{hr^{2n}-r^{2n}_c\left(1-\gamma^{-2}\right)\hat{u}^{2\left(n+2\right)}}\quad.\end{displaymath}
In this form the horizon is apparent, as well as the value of $K$, and the restriction on the zeros of the denominator of $g_{\hat{u}\hat{u}}$ so that the sign change in $g_{\hat{t}\hat{t}}$ is compensated and $-g$ remains positive. It is worth noting that both in this case and for the above extended objects that $\hat{t}$ diverges at the horizon and therefore the hatted coordinates only cover the region exterior to the world-volume horizon, analogous to the Schwarzschild case.

\subsubsection{Energy Loss}
As above, we can fix the sign of $K$ by considering the energy momentum flow. The currents in the radial parameter are:
\begin{eqnarray}\label{eq:eloss}\Pi^{(u)}_t=T_0 R^{n+2}A v^2 K=\pm T_0 R^{n+2}A(1-\gamma^{-2})\gamma^\frac{2\left(n+2\right)}{d}r_c^n\quad,\\
\Pi^{(u)}_x=-\Pi^{(u)}_t/v\quad.\end{eqnarray}
We see, in order for $-\Pi^{(u)}_t$ to be positive, so that energy only flows into the black hole, we demand that $K<0$. Interestingly, the defect only experiences energy and momentum loss if it has some non-zero $r_c$. We will come back to this after examining some solutions.

\subsection{Numerical Solutions}
\subsubsection{Straight Solution}
We first examine the case of a constant profile in the direction of motion, $x$. This corresponds to $x_{,u}$ and therefore $r_c$ being zero. This causes $g_{\hat{u}\hat{u}}$ to be everywhere positive: therefore the defect cannot extend past the critical radius $u_c$ as there is no way to compensate the change in sign of $g_{\hat{t}\hat{t}}$ and maintain the signature of the induced metric. We therefore look for solutions that ``cap off'' smoothly at some $u_0<u_c$. Given the spherical symmetry these will look something like a bowl with opening on the boundary. 

\begin{figure}[!htb]
	\centering
	\includegraphics[width=0.4\textwidth]{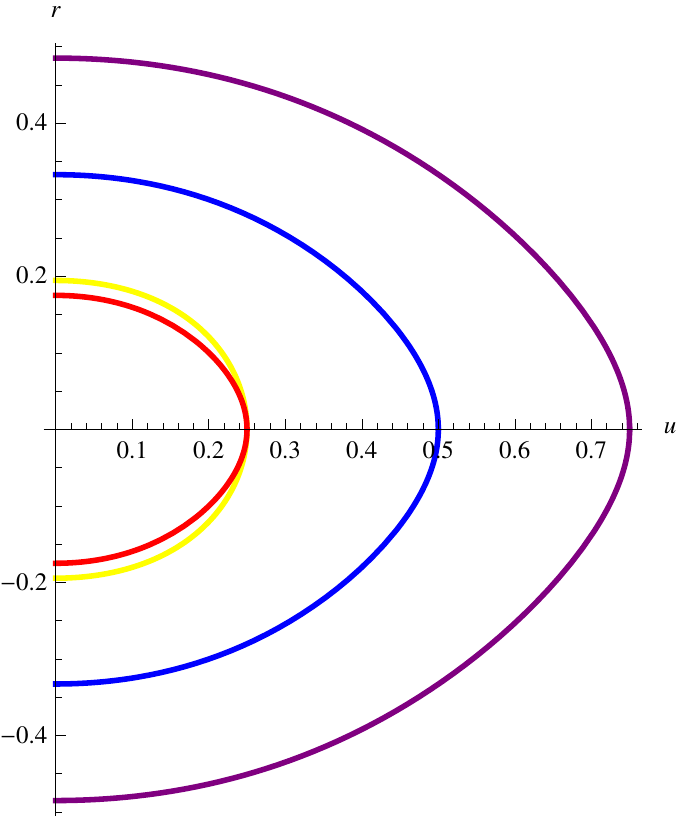}
	\caption{Bowl solutions of the $r$ profile for n=1, d=6. From left to right: Red- $\gamma=40$, $u_0=1/4$; Yellow- $\gamma=2$, $u_0=1/4$;  Blue- $\gamma=6$, $u_0=1/2$;  Purple- $\gamma=2$, $u_0=3/4$;}
	\label{fig:fig4}
\end{figure}

We find solutions by numerically integrating the equation of motion (\ref{eq:eomr}) with the initial conditions $r(u_0)\to 0$ and $r'(u_0)\to -\infty$, where the second condition assures us of smooth capping off, eliminating solutions with a conical singularity. Profiles of defects with $n=1$ in $AdS_7$, for various $\gamma$ and $u_0$, are shown in Figure(\ref{fig:fig4}). As these solutions do not extend past the critical radial coordinate $u_c$ they have $r_c=0$ and from (\ref{eq:eloss}) they experience no energy nor momentum loss.

\begin{figure}[!htb]
	\centering
	\includegraphics[width=0.62\textwidth]{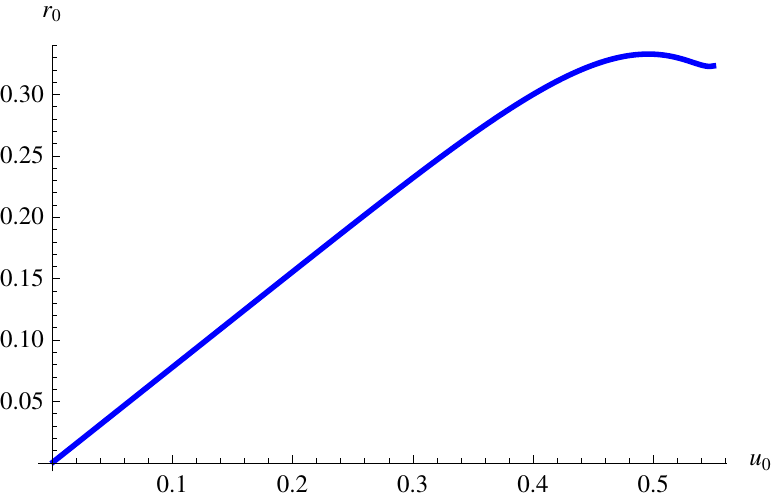}
	\caption{Opening radius of the bowl solution 
on the boundary for n=1, d=6, $\gamma=6$. Note the bowl solution only exists for $u_0<u_c=6^{-1/3}\approx0.55$.}
	\label{fig:fig5}
\end{figure}

For a fixed $\gamma$, as we vary $u_0$, it is interesting to examine the behavior of $r_0$, the radius of the opening of the bowl on the boundary. Of consequence is that $r_0$ attains a maximum for some $u_0<u_c$, as seen in Figure(\ref{fig:fig5}). Alternatively, we can state this as the fact that for a given $\gamma$ there is a maximal radius of the sphere on the boundary for which the bowl solution exists. We will return to the implications of this after examining other solutions.

\subsubsection{Curved Solution}
\begin{figure}[!htb]
	\centering
	\includegraphics[width=0.8\textwidth]{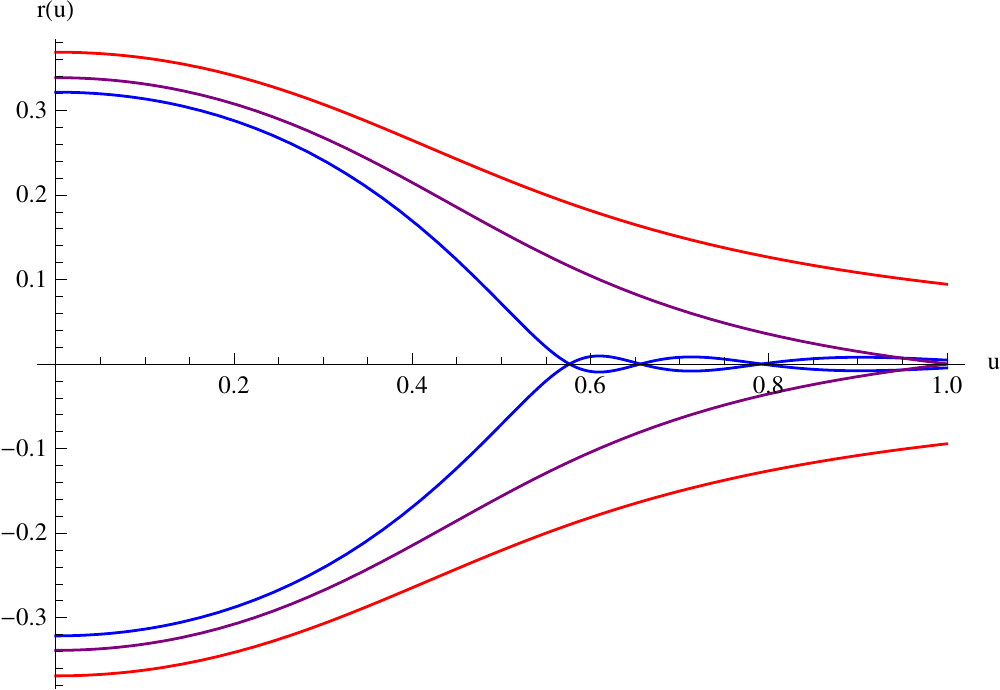}
	\caption{Tube solutions, the $r$ profile for n=1, d=6, $\gamma=6$ and, from inner to outer: Blue- $r_c=0.020$; Purple- $r_c=0.129$; Red- $r_c=0.200$.}
	\label{fig:fig6}
\end{figure}

To have a profile that extends beyond the world-volume horizon at $u_c=\gamma^{-\frac{2}{d}}$ we expect a non-zero $x_{,u}$, as in Section 2. Indeed we see that this is a requirement, so that $-g$ as given by (\ref{eq:detg}) remains positive. Numerically, we look for solutions in the two regions, near the boundary with $0<u<u_c$, and near the black hole with $u_c<u<1$, and match them to a local series solution at the world-volume horizon $u_c$. For example, since near the boundary $u^d<u_c^d=\gamma^{-2}$ we have $(\gamma^{-2}-u^d)>0$ in equation (\ref{eq:detg}) for $-g$, and therefore require that the denominator, $hr^{2n}(u)-r_c^{2n}(1-\gamma^{-2})\gamma^\frac{4\left(n+2\right)}{d}u^{2\left(n+2\right)}$, is greater than zero in this region. 
Near the world-volume horizon we demand that $r(u)$ has a regular power series expansion. Indeed, writing $r(u)=r_c+r_1(u-u_c)+r_2(u-u_c)^2+\cdots$ we find that $r_1$, and subsequently all higher coefficients, are uniquely fixed
in terms of $r_c$ and $\gamma$ by the equations of motion. We now can numerically integrate the equation of motion (\ref{eq:eomr}) in this region with the initial conditions $r(u_c)\to r_c$ and $r'(u_c)\to r_1$. Repeating in the region near the black hole we obtain ``tube''-like solutions over the full space. For $n=1$ in $AdS_7$ profiles of $r(u)$ for various $r_c$ are shown in Figure (\ref{fig:fig6}). The corresponding profiles of $x(u)$ are qualitatively similar to the extended defects of the previous section and are omitted for brevity.

For a fixed $\gamma$, as we vary $r_c$ for these solutions we see that there is a minimal boundary opening $r_0$ that can correspond to a curved tube configuration. From (\ref{eq:eloss}) we see that these tube solutions, having non-zero $r_c$ will experience energy and momentum loss. 

Generically, for small $r_c$ the profile near the black hole has conical singularities where $r(u)\to 0$, see Figure (\ref{fig:fig6}). None of these are seen to cap off smoothly, like the bowl solution does. From the view of the world-volume these singularities are behind the induced horizon at $u_c$ and therefore, by causality, should not influence physics confined to the defect, such as the momentum currents. Indeed, examining these densities (\ref{eq:eloss}) we see they are determined only by the radius of the tube at the induced horizon, independent of further details of the profile. Therefore the classical properties of the world-volume black hole are unaffected by the presence of the singularities.

As the energy flux $-\Pi^{(u)}_t$ is a constant along the tube, regions where $r(u)\to 0$ have arbitrarily large energy densities. Additionally, for $r\ll R$ the tube is probing sub-Planckian scales. For both these reasons we would expect the classical probe approximation to break down.
Fluctuations around the world-volume, as well as the classical back reaction,
will become important close to the conical singularities. Typically, such singular configurations would be thrown out as irrelevant, 
but as we will see in Figure (\ref{fig:fig7}) there is a range of parameters for which such singular solutions are the only configurations that exist. This means that, although we should not trust the details, there is a range of opening radii $r_0$ and velocities $\gamma$ where there is no classical 
probe description and we must examine quantum corrections and effects of the classical back reaction. The $AdS/CFT$ dictionary for this case gives $R/l_p = 2\left(\pi N\right)^{1/3}$, where $l_p$ is the eleven dimensional Planck length. Corrections around the classical world-volume by quantum fluctuations are suppressed by $1/(T_0 R^3) \sim 1/N$, whereas the classical back reaction on the geometry is suppressed by $(T_0 R^3)/(R M_{Pl,7})^5 \sim 1/N$ (as the relation between the $AdS_7$ and 11d Planck masses is $M_{pl,7}^5 \sim M_p^9 R^4$, due to compactification on the internal $S^4$). We can conclude that there is a range of parameters in the $CFT_6$ for which the proper description of spherical defects requires $1/N$ corrections. 

As a graviton can propagate through the bulk from one of these ``hidden'' singular points, and either rejoin the defect on the other side of the world-volume horizon or reach the boundary, we see that the profiles of energy and momentum density at the boundary, as well as the subleading $1/N$ corrections to the loss rates, will be sensitive to the resolution of these conic singularities from either world-volume fluctuations
or classical back reaction. Much of this section is similar to the scenario of \cite{FM}, although the defects there are asymptotically planar, instead of intersecting the boundary as above. The appearance of regions in the phase diagram where conical singularities are unavoidable is in parallel with a similar phenomenon that was seen in the study of holographic flavor in the presence of electric fields \cite{EMS,AFJK}.

Additionally, some of the tubes have proper radius of order and smaller than the $AdS$ radius of curvature $R$. In analogy with black strings in this regime in $AdS$, we expect these configuration might display an instability of the Gregory-Laflamme type \cite{GL,RG}. Stability analysis of these configurations would elucidate this picture.

\subsubsection{Membrane Hawking-Page Transition in $AdS_7$}
\begin{figure}[!htb]
	\centering
	\includegraphics[width=.95\textwidth]{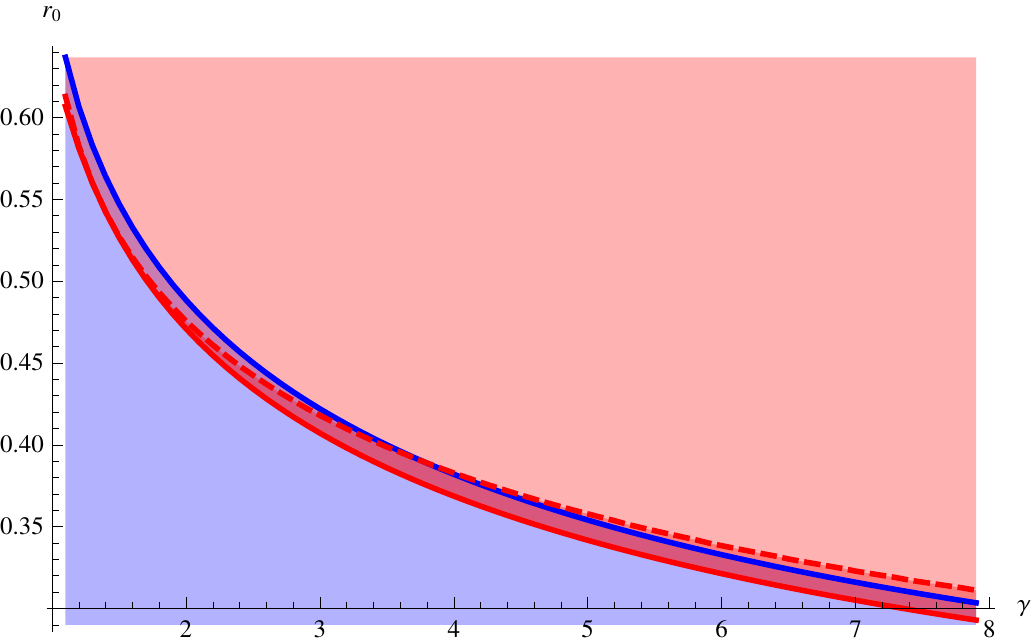}
	\caption{Bowl/Tube phase diagram. Below the (upper) solid Blue line bowl configurations can exist. Above the (lower) solid Red line tube solutions can exist, while between the solid and dashed Red lines these are the singular conic tubes. Note the thin wedge between the dashed Red line and the Blue line where, for $\gamma>3.76$, only the singular solutions exist.}
	\label{fig:fig7}
\end{figure}

In general, for a sphere on the boundary of a given radius $r(0)\equiv r_0$ there could be solutions of either the bowl or tube type. From above, we see that for a given $\gamma$ there is a maximal $r_0$ such that a bowl solution can physically exist, while there is a minimal $r_0$ such that a tube solution is acceptable. Figure (\ref{fig:fig7}) shows how these boundary radial values change for various $\gamma$, for $n=1$ in $AdS_7$.  

A comparison of the free energies of the two configurations in regimes of equal validity is difficult because the tubes experience energy loss while the bowls do not. This means that there must be a constant force on the boundary doing work on the tubes to maintain a constant velocity, and determining the free energy from the classical action becomes subtle. Some progress has been made in this direction by considering the equivalent situation of a defect in the background of a boosted black hole. Regardless, as there are regions of the phase diagram where only solutions of one type can exist, there will be a transition between them. 

This transition can be understood to happen in two different ways. If we consider starting with a tube solution at a speed given by $\gamma$ and a fixed boundary radius $r_0$, we know from (\ref{eq:eloss}) that this configuration will lose energy and momentum. Therefore the defect will tend to slow down, moving to the left in Figure (\ref{fig:fig7}). Eventually it will reach the region where the tube solution is no longer stable, and the configuration will transition to a bowl defect, where upon it will no longer experience energy loss. Alternatively, we can consider starting with a tube solution where we fix the speed given by $\gamma$ and the dimensionful boundary radius $\bar{r}\equiv \frac{d}{4 \pi T}r$. Lowering the temperature of the system now corresponds to decreasing the dimensionless radius $r$. This causes us to move downwards on the figure, until again we reach a region where the tube solution is no longer stable and we transition to a bowl configuration with no energy loss.

From a world-volume view point, this transition is between a spacetime with a black hole (the tube), and one which is horizon free (the bowl.) For this reason we call it a membrane Hawking-Page transition \cite{HP}. It is also very similar to the scenario of ``black hole mergers,'' as discussed in \cite{VF}.

\section{Accelerating Defect}
\subsection{Equation of Motion}
Lastly we look at a solution in a zero temperature $CFT$, i.e. in $AdS$ without a black hole. For this we start with the metric (\ref{eq:metric1}) with $r_h=0$, but make the coordinate change $u=1/r$ so we end up with:
\begin{displaymath} ds^2=\left(\frac{R}{u}\right)^2\left(-dt^2+du^2+d\vec{x}^2\right)\quad.\end{displaymath}
For a defect that, like in the last section, is spatially an $n$-sphere on the boundary, but now has a non static radius $r(t,u)$, we have:
\begin{displaymath} -g=\left(\frac{R}{u}\right)^{2\left(n+2\right)}r^{2n}A^2\left(1+r^2_{,u}-r^2_{,t}\right)\quad.\end{displaymath}
This gives the equation of motion:
\begin{equation}\label{eq:eomacc}\left(\frac{r_{,t}r^n}{u^{n+2}\sqrt{1+r^2_{,u}-r^2_{,t}}}\right)_{,t}=\left(\frac{r_{,u}r^n}{u^{n+2}\sqrt{1+r^2_{,u}-r^2_{,t}}}\right)_{,u}-\frac{nr^{n-1}}{u^{n+2}}\sqrt{1+r^2_{,u}-r^2_{,t}}\quad.\end{equation}

\subsection{Solution}
Inspired by \cite{BWX} we make the ansatz $r(t,u)=\sqrt{t^2+b^2-u^2}$ which indeed satisfies the equation of motion (\ref{eq:eomacc}). This solution describes a bowl-like configuration. The constant $b$ arises as an integration constant, and is seen to be the initial radius of the opening of the bowl on the boundary. Another physical interpretation will be given shortly. In the $u$-$r$ plane this solution is a half circle, with time dependent radius $b^2+t^2$. The bottom of the bowl, where $r=0$, is located at the bulk radial coordinate $u_0^2 = t^2+b^2$. Another important bulk radius, where there will be shown to be a horizon, is $u_b=b$. As we will show in the next section this configuration contains momentum fluxes at the boundary, so it should properly be understood that the total action contains a boundary force term that is countering this energy loss and providing the acceleration of the opening radius. 

\subsection{Energy Loss}
We can again calculate the conjugate momentum densities:
\begin{eqnarray}
\Pi^{(t)}_t=-T_0 A\left(\frac{R}{u}\right)^{n+2}\frac{(t^2+b^2)(t^2+b^2-u^2)^{\frac{n-1}{2}}}{b}\quad,\nonumber \\
\Pi^{(u)}_t=-T_0 A R^{n+2}\frac{t(t^2+b^2-u^2)^\frac{n-1}{2}}{bu^{n+1}}\quad.\nonumber
\end{eqnarray}
Integrating the energy density $-\Pi^{(t)}_t$ from some bulk radius $u$ to the bottom of the bowl at $u_0=\sqrt{t^2+b^2}$ we get the energy in this part of the configuration:
\begin{displaymath} 
E(t,u)=\int_{u}^{u_0}du'\int d\phi\Pi_i d\theta_i\left(-\Pi^{(t)}_t\right)=\frac{T_0 S_nR^{n+2}(t^2+b^2-u^2)^\frac{n+1}{2}}{(n+1)bu^{n+1}}\quad,
\end{displaymath}
where $S_n$ is the area of the unit n-sphere. We note that to have a defect with finite energy we must truncate the configuration at some $u_m>0$, analogous to the case of the string in \cite{HKKKY}. 

Despite the fact that $\partial{r}/\partial{t}$ becomes greater than 1 at bulk radius $u>b$, this is just a coordinate artifact, and to determine the physical acceptability of this configuration we should examine the speed of energy flow. We can consider, as in \cite{BWX},  how fast we have to move the hypersurface $u=u_*$, as energy flows through it, so that $E(t,u_*)=const$, or note that the speed of energy flow, $s$, will be given by the ratio of the rate of energy loss to the energy density: 
\begin{displaymath} s\equiv \frac{-\Pi^{(u)}_t}{-\Pi^{(t)}_t}=\frac{ut}{t^2+b^2}\quad.\end{displaymath}
As on the defect $u\le\sqrt{t^2+b^2}$ we see that this physical speed is always less than that of light. None-the-less, the coordinate $u_b=b$ plays a special role to which we now turn.

\subsection{World Volume Horizon}
Following \cite{BWX} we define a new time coordinate, $\tau$, by $t=\sqrt{b^2-u^2} \sinh\left(\tau/b\right)$. Due to the root, these coordinates only cover the region with $u<b$. This reparametrization gives the diagonal world-volume metric:
\begin{displaymath}
	ds^2_{wv}=\left(\frac{R}{u}\right)^2\left(-\left(1-\frac{u^2}{b^2}\right)d\tau^2+\frac{du^2}{\left(1-\frac{u^2}{b^2}\right)}+\left(b^2-u^2\right)\cosh^2\left(\frac{\tau}{b}\right)d\Omega_n^2\right)\quad.
\end{displaymath}

In this form it is apparent that there is a horizon at $u=b$. Examining the Euclidean metric we see that near this horizon the time and angular variables take the form of an $n+1$-sphere. To avoid a conical defect we require the Euclidean time to be treated as an angular variable with period $2\pi b$, thus giving the temperature of the Lorentzian horizon, $T=1/2\pi b$. Interestingly, even though we started with zero temperature $AdS$, due to the accelerating nature of the defect a world-volume temperature emerges. The existence of energy loss is also directly connected to the fact that there is a world volume horizon, as discussed in \cite{CG}.

We can make an additional coordinate change to illuminate the nature of this world-volume ``cosmology.'' Changing to the tortoise-like radial coordinate $\rho \equiv b \tanh^{-1}(u/b)$, so that the boundary is at $\rho=0$ and the horizon at $\rho=\infty$, the metric becomes:

\begin{displaymath}
	ds^2=\left(\frac{R}{b}\right)^2\frac{1}{\sinh^2\left(\rho/b\right)}\left(-d\tau^2+d\rho^2+b^2\cosh^2\left(\tau/b\right)d\Omega_n^2\right)\quad.
\end{displaymath}
From here we see that the world-volume is conformal to a hyper-cylinder: $\rho$ parametrizes its length, and the n-sphere has an exponentially growing radius. This is to contrast the $n=0$ case of the string discussed in \cite{BWX}, there the n-sphere factor was absent and the world-volume was static. This gives us a novel cosmology to study in the context of $AdS/CFT$. 

For the portion of the world volume below the horizon at $u=b$ we can also diagonalize the metric. It is convenient to first replace the bulk radial variable $u$ with a parameter that is the rescaled radius of the defect: $\hat{r}\equiv r/r(u_b)=\sqrt{1+\left(b^2-u^2\right)/t^2}$, where the bottom of the bowl is at $\hat{r}=0$ and the horizon is at $\hat{r}=1$, for all $t$. Then, defining: 
\begin{displaymath}
	\tau\equiv \tan^{-1}\left(\frac{t\sqrt{1-\hat{r}^2}}{b}\right)\quad,\quad w\equiv \frac{1}{2}\log\left(\frac{1+\hat{r}}{1-\hat{r}}\right)\quad,
\end{displaymath}
the world-volume metric becomes:
\begin{displaymath}
	ds_{wv}^2=R^2\left(-d\tau^2+\sin^2(t)\left(dw^2+\sinh^2(w)d\Omega_n^2\right)\right)\quad,
\end{displaymath}
with $w=0$ at the bottom of the bowl, $w=\infty$ at the horizon, and $\tau$ ranging from $0$ to $\pi/2$. In this form it is seen that the world-volume cosmology is that of an open FRW universe, with scale factor $a(\tau)=\sin(\tau)$. There is a ``big bang'' at $\tau\to 0$ where $a\to 0$, and then the scale factor grows monotonically until the ``end of the universe'' at $\tau=\pi/2$. The fact that the proper age of the world-volume universe is finite, namely $T=\pi R/2$, is understood by noting that the geometry is just that of $AdS_{n+2}$ written in a hyperbolic slicing. The above coordinates cover the inside of a light cone starting at a point and ending on the boundary, in essence the compliment of the region considered in \cite{RE}.

\section{Conclusion}
We have studied probe defects moving through Anti-de Sitter space. The major lesson learned is the importance of world-volume horizons in determining the physical behavior. We see that observables like the energy loss of the moving defect are determined by the effective temperature of this induced horizon, not simply the temperature of the bulk black hole, which corresponds to that of the CFT plasma. This leads us to conclude that the probe defect only interacts with a blue shifted energy density, and that the hydrodynamic description remains accurate for these moving objects, consistent with but signficantly extending earlier studies. 

The main example of focus, that of a membrane in $AdS_7$, has been a worthwhile study. The special analytic properties of this situation have been useful in determining the profiles for the infinitely extended defects. Despite having to resort to numerics for compact defects, we stuck with this example, as we still expect it to be special based on its underlying M-Theory description. 

As a first step in further study of these scenarios, we have previously done the stability analysis for the infinitely extended defects. Using a generalized continued fraction technique we calculated the first few quasinormal mode frequencies. There was no appearance of modes with positive imaginary frequency, as would be a sign of instability. Interestingly, a massless Nambu-Goldstone mode, due to the translational symmetry breaking of the defect, was seen analytically and deserves further study. A stability analysis of the compact solutions is desired. This would add clarity to the details of the phase transition between bowls and tubes.   

The effects of quantum gravity, $1/N$ corrections should also be explored. In line with our general theme, it would be interesting to see how these corrections effect things like energy loss, as they allow communication from parts of the world-volume behind the induced horizon with the rest of the defect. For the case of compact defects we have the further issue that these horizons hid conical singularities in the profile. How corrections smooth out these regions and effect the above description would add clarity to the physicality of these solutions.  

\section*{Acknowledgments}
This work was supported in part by the U.S. Department of Energy 
under Grant No. DE-FG02-96ER40956.

\bibliography{movingdefects}{}
\bibliographystyle{elsart-num}

\end{document}